\documentclass[12pt]{article}

\usepackage{scicite}

\usepackage{times}
\usepackage{graphicx}

\usepackage{placeins}

\newenvironment{sciabstract}{%
\begin{quote} \bf}
{\end{quote}}

\newcounter{lastnote}

\title{Breaking Cascadia's Silence: Machine Learning Reveals the Constant Chatter of the Megathrust}

\author
{Authors: Bertrand Rouet-Leduc$^{1\ast}$, Claudia Hulbert$^{1\ast}$, Paul A. Johnson$^{1}$
\\
\normalsize{$^{1}$Affiliation: Los Alamos National Laboratory, Geophysics Group, Los Alamos, New Mexico, USA}\\
\\
\normalsize{$^\ast$ These authors contributed equally to the work}\\
\normalsize{B. Rouet-Leduc (bertrandrl@lanl.gov), C. Hulbert (chulbert@lanl.gov) and P. Johnson (paj@lanl.gov)}
}

\date{}

\begin{document} 


\baselineskip24pt


\maketitle

\subsection*{Short Title}
Breaking Cascadia's Silence

\subsection*{One Sentence Summary}

Continuous seismic signals emanating from the Cascadia subduction zone and recorded on Vancouver Island, Canada are imprinted with the instantaneous subducting fault slow-slip, at all times during the slow slip cycle.

\subsection*{Abstract}
\begin{sciabstract}

%Tectonic faults slip in various manners, ranging from ordinary earthquakes to slow slip events to aseismic fault creep. Slow earthquakes and associated intermittent tremor bursts, in particular, are poorly understood. We show that the Cascadia megathrust is continuously broadcasting a tremor-like signal that precisely informs of fault displacement rate throughout the slow slip cycle, that may account for much of the discrepancy between surface displacement and cumulative seismic tremor energy. We posit that this tremor-like signal provides indirect real-time access to friction on the Cascadia megathrust, and thus may prove useful in determining if and how a slow slip may evolve into a megaquake in neighboring regions of the seismogenic zone.

Tectonic faults slip in various manners, ranging from ordinary earthquakes to slow slip events to aseismic fault creep. The frequent occurrence of slow earthquakes and their sensitivity to stress make them a promising probe of the neighboring locked zone where megaquakes take place. This relationship, however, remains poorly understood. We show that the Cascadia megathrust is continuously broadcasting a tremor-like signal that precisely informs of fault displacement rate throughout the slow slip cycle. We posit that this signal provides indirect, real-time access to physical properties of the megathrust and may ultimately reveal a connection between slow slip and megaquakes.

\end{sciabstract}

\subsection*{Main text}

\begin{figure}[ht!]
\begin{center}
\includegraphics[width=15cm,trim= 0 0 0 0]{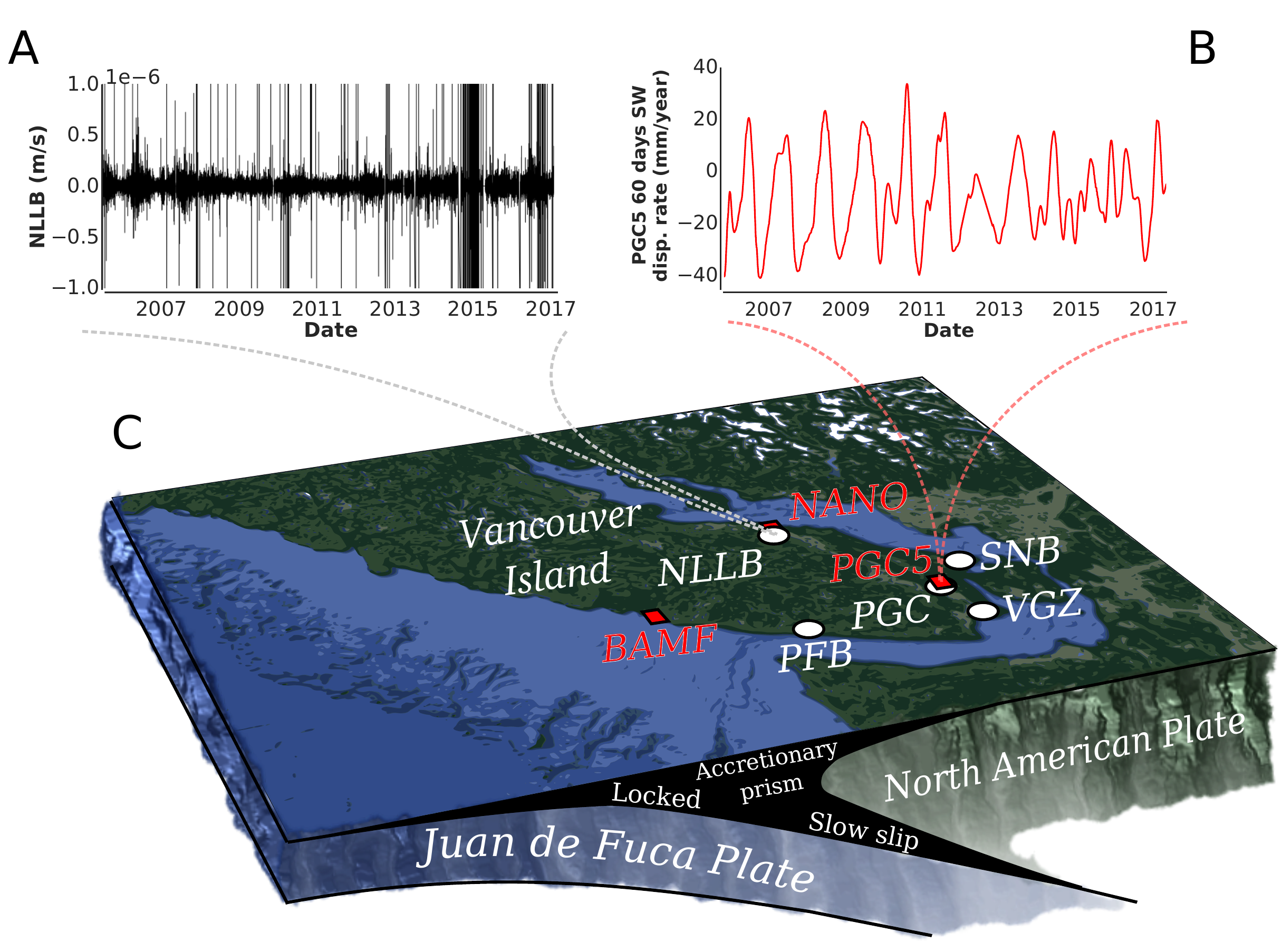}
\caption{\footnotesize{\textbf{Map and schematic of the region analyzed: Vancouver Island and the subduction zone.} Our goal is to determine if, as in the laboratory, characteristics of the full continuous seismic data (\textbf{A}) can be used to accurately estimate the GPS displacement rate (\textbf{B}). Note that the seismic data shown is clipped to eliminate earthquakes so that waveform structure between slip events used in the analysis can be seen. (\textbf{C}) The seismic stations are shown in white and the GPS stations in red.}}
\label{fig:fig1}
\end{center}
\end{figure}

Following the discovery of tectonic tremor in Japan by Obara in 2002\cite{Obara2002}, Rogers and Dragert identified episodic transient slow slip on the Cascadia fault beneath Vancouver Island, Canada\cite{Rogers2003}. The slow slip occurs roughly every 13 months, when the over-riding North American plate lurches southwesterly over the subducting Juan de Fuca plate (see Fig. \ref{fig:fig1}). 
Subduction zones are divided into regions that slip during earthquakes and regions that slowly slip, mostly aseismically\cite{Perfettini2010}. Slow slip in Cascadia is located down-dip from the seismogenic rupture zone, cyclically loading the locked fault\cite{WechCreagerMelbourne2009,bartlowGRL2011,Kao2006,Kao2009,wech2008} (Fig. \ref{fig:fig1}) that has not broken since \emph{ca.} 1700 \cite{Goldfinger2003}. 
Slow, aseismic slip on the plate interface is accompanied by discrete bursts of tectonic tremor, potentially radiating from strong asperities \cite{WechBartlow,Rogers2003}. Recently, smaller slow slips taking place intermittently between the large, periodic ones have been identified, indicating that slow slip occurs over a spectrum of time scales\cite{Frank2016,HawthorneRubin2013}.

Elastic wave radiation by fracture propagation provides an accurate model of regular earthquakes\cite{Brace1966,Scholz2002}. In contrast, the physics of slow slip and associated  tremor is not well understood\cite{Obara253}. Several questions remain regarding the slip frictional physics, rupture propagation, and the relationship between slip modes \cite{Satake1996,Obara253}.  Slow slip has been observed preceding large subduction earthquakes \cite{ITO201314,hasegawa2015preceding,Kato705,Radiguet2016,Ruiz1165}, suggesting that they may modulate or be part of the nucleation process of major fault ruptures. Studies also find a large discrepancy between slow slip surface displacement and cumulated tremor energy--seismic moment calculated from identified tremor episodes are several orders of magnitude below the estimated energy release inferred from GPS measured displacement\cite{aguiar2009moment,Kao2009,kostoglodov20102006}.
%However, slow earthquakes and huge megathrust earthquakes can have common slip mechanisms and are located in neighboring regions of the seismogenic zone. The frequent occurrence of slow earthquakes may help to reveal the physics underlying megathrust events as useful analogs. Slow earthquakes may function as stress meters because of their high sensitivity to stress changes in the seismogenic zone.

We find that the Cascadia megathrust emits a tremor-like signal apparently at all times, which may account for most of this missing energy. Statistical characteristics of this signal are a fingerprint of the displacement rate of the fault. These results are surprisingly close to our previous work on laboratory slow slip \cite{Hulbert2017SlowSlip}, suggesting that the underlying physics may scale from a laboratory fault to Earth. If this is the case, this continuous signal may give us an indirect measure of the friction on the fault (as observed in the laboratory), and may help unravel fundamental questions about slow earthquakes: what is the mechanism behind the quasi-dynamic slow slip propagation and can large earthquakes be triggered by slow slip events? Slow earthquakes and megaquakes come from neighboring regions of the megathrust and improving our understanding of the former may unravel the physics of the latter.

%A key question for geoscientists is to determine whether episodic tremor and slip provides clues to an anticipated megaquake \cite{Satake1996,Obara253}. including the M9 Tohuku, Japan earthquake \cite{ITO201314,hasegawa2015preceding,Kato705}, the 2013 M7.3 Papanoa earthquake in Guerrero, Mexico\cite{Radiguet2016} and the M8.1 Iquique, Chile earthquake \cite{Ruiz1165}  %As we advance our understanding of slow slip physics and its coupling to the megathrust zone, this relationship may be unraveled, leading to marked improvements in hazards assessment in Cascadia and elsewhere \cite{Satake1996}. \\

%In laboratory experiments of slow slip, we found that the full continuous seismic signal emitted by artificial faults is imprinted with rich information regarding their physical state \cite{Hulbert2017SlowSlip}. In the laboratory, analysis of continuous seismic waves enables us to accurately estimate fault friction and shear displacement, and reveals precursory information regarding the timing and magnitude of upcoming slip events \cite{Hulbert2017SlowSlip,rouet2017fault}.

%In this work we scale the laboratory approach to Earth. We use GPS displacement rate as a proxy for the frictional state of the Cascadia fault. 

\paragraph*{The displacement rate measured by GPS can be accurately estimated from the continuous seismic data.}

Analyses based on earthquake and tremor catalogs offer important insight into slip processes, but discard most  of the continuously recorded seismic data. The approach we describe harnesses the power of supervised learning using other reliable geophysical datasets as a label to extract information from continuous seismic data, in contrast to labels built by hand or heuristics. More specifically, we pose the problem as a  regression between characteristics of the continuous seismic data and the displacement rate of the subduction zone obtained from GPS.

We rely on seismic data from the Canadian National Seismograph Network (CNSN) \cite{FDSN}. In order to measure the displacement rate presumed to be taking place at the fault interface, we use processed data \cite{USGS} from nearby GPS stations from the USGS Pacific Northwest Network. We consider the total horizontal displacement (East/West + North/South GPS components), such that a negative displacement is in the direction of the subduction (North-East), and a positive displacement is in the direction of the slip (South-West). Both the seismic and GPS stations we used in this study are shown in Figure \ref{fig:fig1}. The seismic data are recorded or re-sampled at 40 Hz, and the GPS data are processed at 1 sample per day.

%We begin by processing the seismic and GPS data. For each station and for each day, we correct the seismic signal for the instrument response, remove the average and detrend the data. We make several copies of the seismic signal, bandpassed at frequencies between 8 to 13 Hz, in increments of 1 Hz (8 to 9 Hz, 9 to 10 Hz, 10 to 11 Hz, 11 to 12 Hz, 12 to 13 Hz). While tremor exists down to frequencies of approximately 1 Hz, in the 8-13 Hz band the signal is nonetheless strong, and contributions from microseisms that could affect the analysis are very weak \cite{Zhang2011}. Following  bandpassing, the signal is clipped ($10^{-7}$ m/s) to remove impulsive signals arising from regular earthquakes \cite{Bensen2007}. This is done because we are interested in analyzing the background signal, and by doing so we avoid the contamination of our statistical features by earthquake waveforms (Figure \ref{fig:fig1}). 
The raw GPS data are extremely noisy, and therefore we process them to emphasize the slow slip events. The denoised displacement rate is calculated as the moving average of the slope of a least squares linear regression over the time window $n_{\rm GPS}$ considered.

Once the data are processed, we construct the ML model. We scan the processed seismic data to extract statistical features, including amplitude and frequency characteristics. These features are used as input to the ML model; the output of the model is the GPS displacement rate. 
Statistical features are calculated for hourly/daily segments of the seismic signals. These hourly/daily features are then averaged over the time interval considered. This is accomplished using time windows with fixed duration. These time windows correspond to a fraction of the length of the slow slip cycle which is approximately 13 months. We find that the window length $n_{\rm{seismic}}$ has little impact on our results (we varied the window from 1 hour to 60 days--see Fig. 1 and 2 of the Supplementary). Features averaged over $n_{\rm seismic}$ hours/days are used to estimate the average displacement rate over $n_{\rm GPS}$ days, with the first day of both windows coinciding. We index the figures and our ML database with the last hour/day of the GPS window.
In the following, `seismic features' will refer to  features averaged over a time window $n_{\rm seismic}$. We know from our previous studies of laboratory shear experiments that distribution-related features are the most important to probe fault characteristics.\\

Once all the seismic features are calculated for all the seismic stations, they are used as inputs to build our ML model. As in any supervised machine learning approach, the model is created in the training phase, for which the algorithm has access to both the features of the continuous seismic data and to the GPS data (training set), and attempts to uncover the underlying function relating these seismic features to GPS displacement rates - thus formulating the problem as a regression. 
The model is then evaluated on a new dataset the model has never seen, the testing set, where it has access only to seismic features. For the testing set, the GPS displacement data are only used to evaluate the performance of the model, \emph{i.e.} the agreement between the model estimates and the true GPS displacement values. We use the first 4 years of the seismic features and GPS data as the training set (from 2005 to 2008), and the following 8 years (from 2009 to 2017) as the testing set. The model's performance in testing is evaluated by using the Pearson correlation coefficient, a standard regression metric that measures the linear correlation between two variables. We rely on Bayesian optimization to set the model's hyperparameters, by 5-fold cross-validation. 

\begin{figure}[ht!]
\begin{center}
\includegraphics[width=15cm,trim= 0 0 0 25]{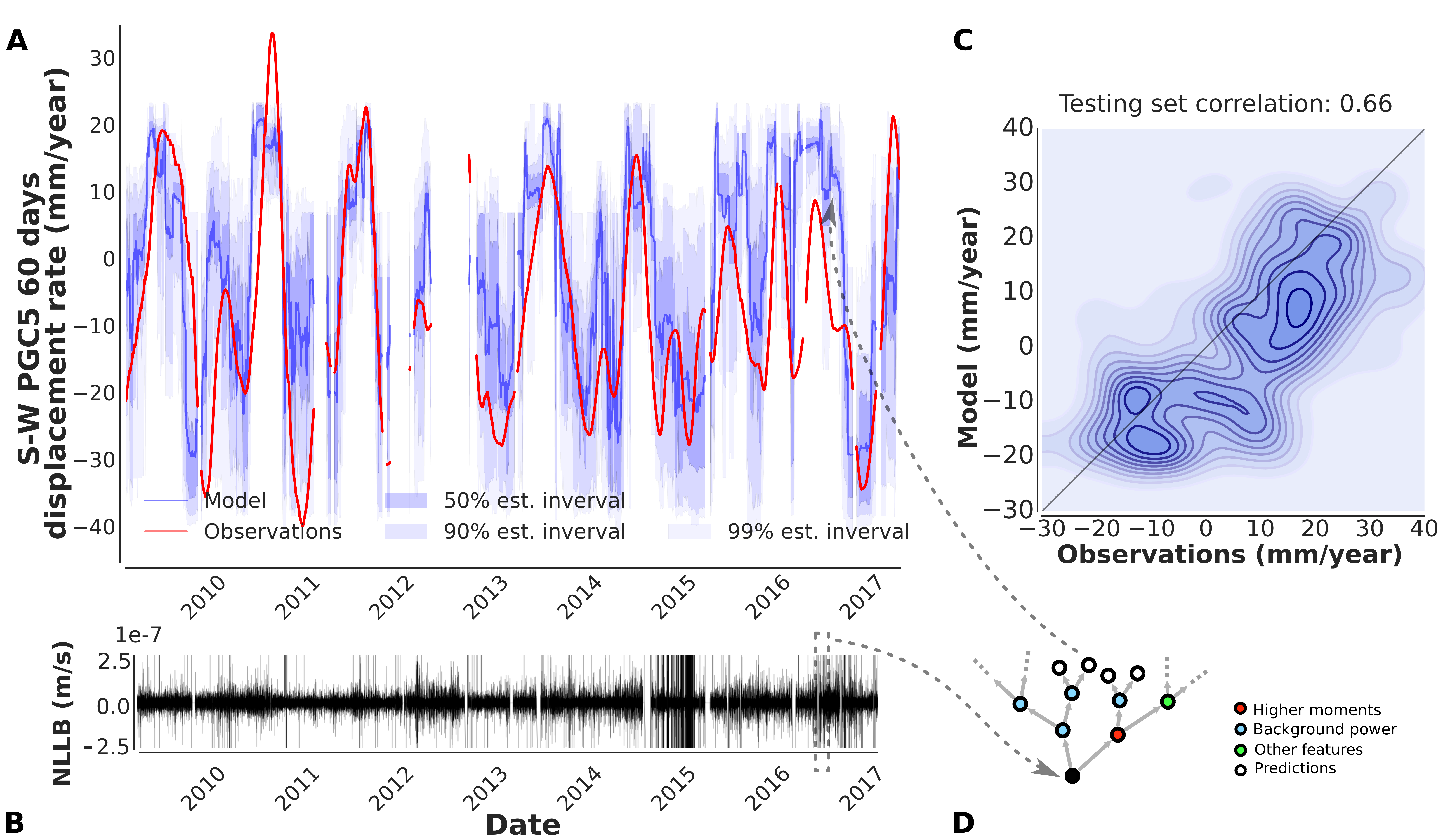}
\caption{\footnotesize{\textbf{Estimating the GPS displacement rate from the continuous seismic data.}  (\textbf{A}) The red line shows the actual, smoothed GPS displacement rate (60 day, from the PGC5 GPS station). The blue bold curve is an estimate from the ML model (with estimation intervals noted with shades of blue) using characteristics of the full continuous seismic data as input. The figure shows the testing set, for which the algorithm \textit{only} has access to the seismic data (e.g., (B)). The data gaps indicate missing (GPS and/or seismic) data.  (\textbf{B}) Continuous seismic data from station NLLB over the same time interval. (\textbf{C}) Distribution of observed versus predicted displacement rates, with contours showing empirical iso-density, from 10 to 90$\%$. The Pearson correlation coefficient between estimates and actual displacement rate is of 0.66, showing that continuous seismic waves contain rich information about the fault's state, apparently at all times. (\textbf{D}) Once the full continuous seismic data has been turned into a database of statistical features, it is fed to an ensemble of decision trees (schematic) that partitions the data to build a model of GPS displacement rate as a function of statistics of the seismic data. }}
\label{fig:fig3}
\end{center}
\end{figure}

Fig. \ref{fig:fig3} shows the model estimations of displacement rate on the testing set for the GPS station PGC5 as a blue bold line, for $n_{\rm seismic}=n_{\rm GPS}=60$ days. The actual GPS-derived displacement rate is shown in red.  

%and for $n_{\rm seismic}=1$ day, $n_{\rm GPS}=30$ days (therefore estimating the displacement rate over the future 30 days from one day of seismic data). 

We also build models estimating the GPS displacement rate from the seismic data for the other GPS stations (see Fig. 3 of the Supplementary). Assuming the GPS data are a proxy for the displacement rate of the plate-interface, these results show that continuous seismic waves can be used to directly estimate the fault's displacement rate at different locations on Vancouver Island. The GPS displacement rates estimated from seismic waves are highly accurate, with Pearson correlation coefficient between estimated and real values above 0.6 for $n_{\rm seismic}=n_{\rm GPS}=60$ days, above 0.5 for $n_{\rm seismic}=n_{\rm GPS}=30$, above 0.4 for $n_{\rm seismic}=1$ day and $n_{\rm GPS}=30$ days, and close to 0.4 for $n_{\rm seismic}=1$ hour, $n_{\rm GPS}=60$ days (see Fig. 1 and 2 of the Supplementary for models using shorter windows than 60 days). These results show that continuous seismic waves contain rich information about the fault's displacement rate on an \textit{hourly} basis. In the Supplementary we show results for a wide range of seismic and GPS windows.

%The ML model estimates are very accurate except early in 2015.  We hypothesize that the slip mechanism for this period may be different, and is therefore missed by our model.  This is the case at all stations and window sizes. Unusually large earthquakes occurred in the area shortly before this time, including an M4.8 earthquake on January 8, 2015 near Tofino, Canada, and may have been coupled with a change in slow slip behavior.  

%%%%%%%%%%%%%%%%%%%%%%%%%%%%%%%%%%

\paragraph*{The Cascadia subduction zone continuously broadcasts signatures of its displacement rate.} 

The random forest machine learning algorithm we use\cite{Breiman1999} is transparent and easy to probe for physical information, in contrast to more `black-box' approaches such as deep neural networks. The model relies on an ensemble of decision trees to directly and explicitly build an estimator of GPS displacement rate from features of the seismic data. This allows us to probe the model in order to identify which features are critical to estimate the fault's displacement (Figure \ref{fig:fig3}). 

When analyzing earthquake cycles from a laboratory fault, one can derive a similar ML model relating fault displacement (or fault friction) to features of the continuous seismic signal emitted from the fault. In the case of the laboratory, the most important feature is the power of the seismic signal. In the laboratory, determining the power of the seismic signal at any given moment is sufficient to accurately infer the frictional state of the fault at that same moment, leading to an equation of state between seismic power and friction \cite{Rouet2017}.

\begin{figure}[ht!]
\begin{center}
\includegraphics[width=15.5cm,trim= 0 0 0 0]{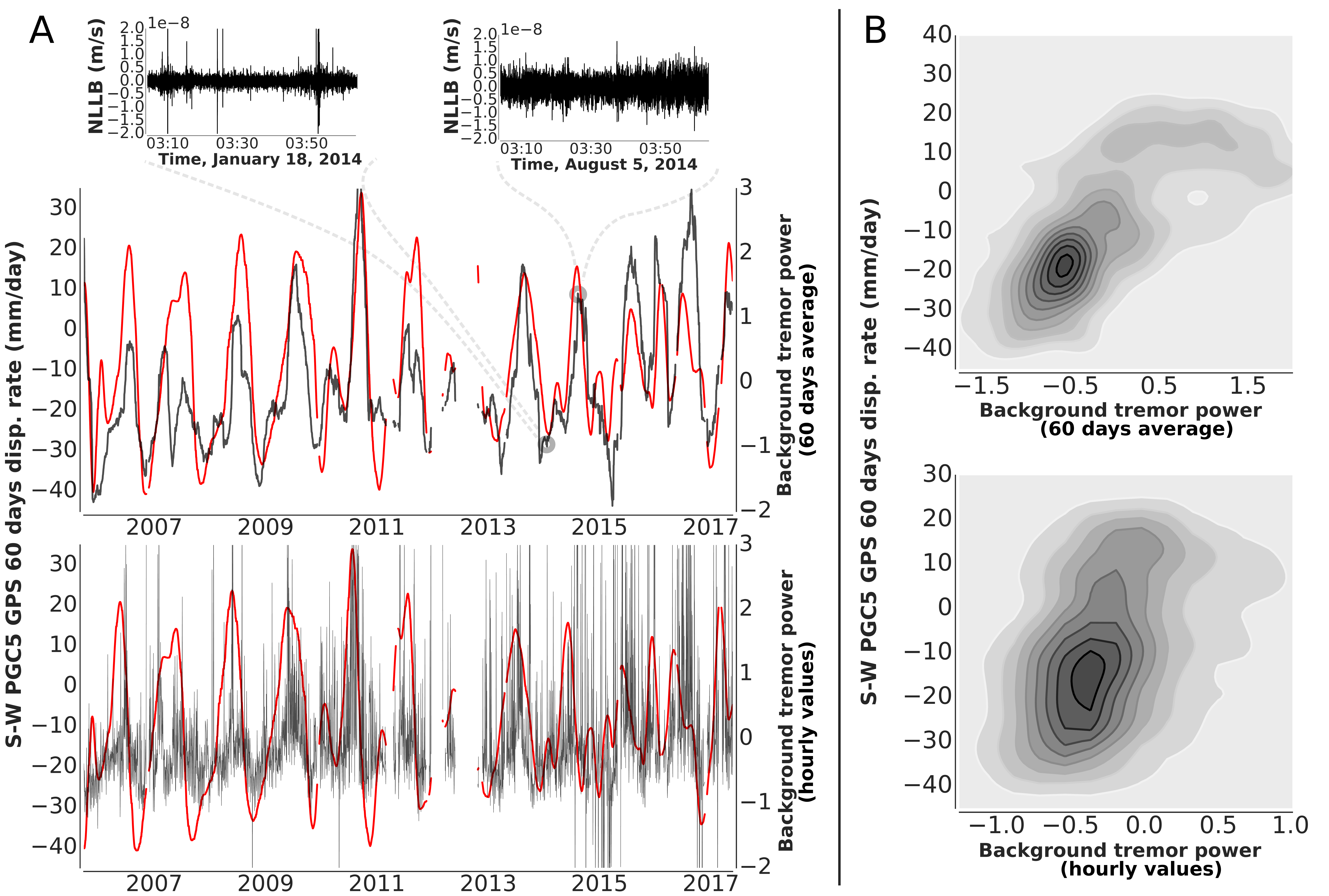}
\caption{\footnotesize{\textbf{The machine learning model uncovers strong correlations between features of the seismic data and the displacement rate of the fault.}
(\textbf{A}) The smoothed displacement rate at the PGC5 GPS station is in red, the best feature of the seismic data (as identified by the RF model) is in black, averaged over 60 days (upper plot) or one hour (lower plot).
The upper insets show seismic data when the statistical feature is low [left] versus when it is high [right], illustrating the changes in background tremor power captured by this feature.  It is this change over the fault cycle that leads to accurate slip-rate estimates.
(\textbf{B}) Density plot showing the empirical distribution of displacement rate at the PGC5 GPS station versus the two best features of the seismic data, for all the data considered (from 2005 to 2017). The $y$ axis is the average displacement rate within a window of 60 days, and the $x$ axis is the corresponding normalized value of the feature, during the same window for the upper plot, from hourly values at the start of the GPS window for the lower plot. Gaps indicate missing data. Slow slips are very clearly identifiable down to an hourly scale in the seismic tremor-like signal, whereas GPS data must be smoothed and averaged over long periods of time to reveal slow slips.}}
\label{fig:fig4}
\end{center}
\end{figure}

In Fig. \ref{fig:fig4} we show the most informative statistical feature of the seismic signal that the random forest identifies and uses to build an accurate model of displacement rate, for $n_{\rm seismic}=n_{\rm GPS}=60$ days, and for $n_{\rm seismic}=1$ hour, $n_{\rm GPS}=60$ days. This feature accounts for most of the fault displacement rate estimates shown in Fig. \ref{fig:fig3}, but Fig. \ref{fig:fig4} simply shows the evolution of this feature and of GPS displacement rate over time, with no machine learning involved.

%<=======================  

The most informative features are proportional to the displacement rate of the fault, increasing whenever the North American plate slips towards the south-west (see references \cite{rouet2017fault,Hulbert2017SlowSlip} and Fig. 4 of the Supplementary for a comparison with lab results). Calculating the tremor `power' over a window of seismic signal enables a precise estimation of the average displacement rate of the plate. This result demonstrates that a statistic of the continuous seismic wave tracks the GPS slip, showing that the accuracy of the random forest model cannot be an artifact.

Interestingly, in the case of the Cascadia slow earthquakes, the features identified as most important by the ML model are very similar to those found in the laboratory (see Fig. 4 of the Supplementary).
Just as in the laboratory, the most important features are related to the power of the signal. The fact that very similar features are found in the laboratory and Earth suggests that the underlying physics scale from a laboratory fault to Earth. 

\begin{figure}[ht!]
\begin{center}
\includegraphics[width=14cm,trim= 0 0 0 0]{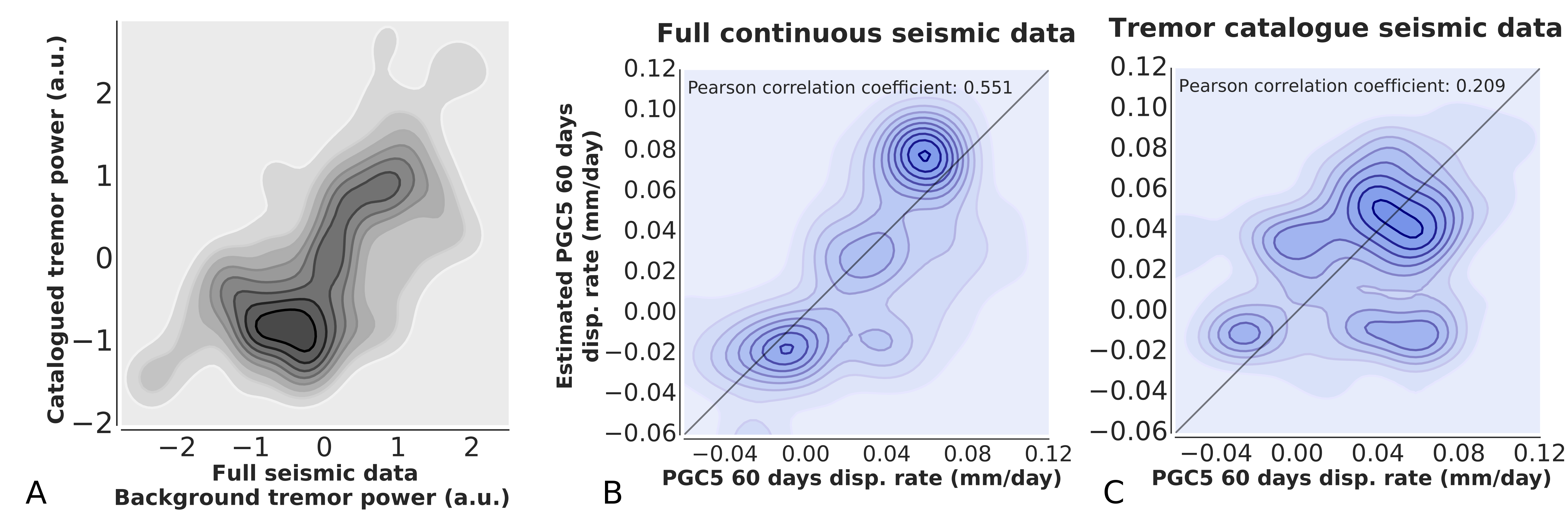}
\caption{\footnotesize{\textbf{The continuous tremor we identify appears to be a continuation of known tremor bursts.} (\textbf{A}) Distribution from 2009 to 2017 of the 60 day average normalized best feature of the seismic data (as identified by the RF model) from the full seismic data ($x$ axis) versus the same normalized feature from the seismic data \textit{only} during catalogued classical tremor events ($y$ axis) from the PNSN tremor catalogue. When the (normalized) background tremor is high ($>0$), corresponding to high slip rates episodes, the classical tremor event power and background tremor power are proportional, suggesting a common origin.  (\textbf{B}) GPS displacement estimation in testing from the full continuous seismic data (same as in Fig. \ref{fig:fig3}, using data from 2009 to 2012 for training and 2013 to 2017 for testing). (\textbf{C}) GPS displacement rate estimation from the continuous seismic data, \emph{restricted} to times of identified tremor events from the PNSN tremor catalog, using the same ML-based method  described in the text to produce the results of (B) here and Fig. \ref{fig:fig3}. Seismic data during the cataloged tremor events are informative of the displacement rate only when at its highest, whereas the remainder of the seismic data are imprinted with information regarding the displacement rate, apparently at all times.}}
\label{fig:fig5}
\end{center}
\end{figure}

Our results suggest that tremor in Cascadia occurs all the time, or nearly all the time. The point is made more clearly in Figure \ref{fig:fig4} where we show that the background tremor energy tracks the GPS displacement rate on an hourly basis. This means there is sufficient tremor in any hourly segment of the  data we analyzed to obtain physically meaningful tremor statistics informing us of the fault slip rate. We note that one could argue that we record only bursts of tremor during each of those hours and that is sufficient to estimate displacement rates. This is possible, and we cannot dispute this point, but at the very least our results show that tremor is much more frequent than previously determined, as has been suggested by other authors \cite{Frank2015,Frank2016,HawthorneRubin2013}. 

It has been established that the accumulated seismic moment of catalogued tremor is several orders of magnitude too small to account for observed slow slip surface displacement \cite{aguiar2009moment,Kao2009,kostoglodov20102006}. A very rough analysis integrating our continuous tremor-like signal over time shows that it contains about 300 times more energy than catalogued tremor events, accounting for much of this missing energy (see Supplementary for details). This signal may originate from local heterogeneities radiating low-amplitude seismic waves within the slipping region \cite{bartlowGRL2011}, modulated by more energetic signals previously identified in tremor catalogs (primarily during peak slow slip) that may be associated with abrupt engagement of large asperities. The strict correlation during peak slow slip between catalogued tremor energy and energy from the tremor-like signal we find (see Fig. \ref{fig:fig5}) suggests a common origin. 

These results also indicate that the slip can occur below the GPS threshold and that the seismic signal may be a more sensitive indicator of slip than GPS, as suggested by others \cite{WechCreagerMelbourne2009,Frank2016}. Our continuous tremor-like signal tracks the state of the slowly slipping fault on an hourly basis, while GPS displacement rates are too noisy to determine fault slip when calculated on windows smaller than a few days (see Fig. 3 here and Fig. 1 of the Supplementary). 

\paragraph*{Conclusions.}
 
The continuous signal we identify tracks the slow slip rate, apparently at all times, providing real time access to the physical state of the slowly slipping portion of the megathrust. As the slow earthquakes transfer stress to the adjacent locked region where megaquakes originate, careful monitoring of this tremor-like signal may provide information on the locked zone, with the potential to improve earthquake hazard assessment in Cascadia. \\

%\pagebreak

\textbf{Acknowledgements}.  This work was funded by Institutional Support (LDRD) at Los Alamos, and by US DOE Fossil Energy.  We are grateful to Herb Dragert, Honn Kao, Tim Melbourne, Joan Gomberg, Daniel Trugman and Ian McBrearty for fruitful comments and discussions. We thank William Frank for his extensive review of our work and his suggestions. We thank Tim Cote, Xiuying Jin and Michal Kolaj from the Canadian National Seismograph Network for their data and their help. PAJ devised the original study. BRL and CH conducted the machine learning work and all authors contributed to writing the manuscript. The authors have no competing interests.   The seismic data used here come from the Canadian National Seismograph Network \cite{FDSN}, and the GPS data come from the Pacific Northwest network of the USGS (doi:10.5066/F7NG4NRK).

\bibliographystyle{ScienceAdvances}

\end{document}